\documentclass[iop]{emulateapj}
\usepackage{amsmath,dsnr,epsfig,hyperref,ifthen,natbib}
\shorttitle{} \shortauthors{Rupke \& Veilleux}

\begin{document}

\journalinfo{The Astrophysical Journal Letters, 729:L27, 2011 March
  10}

\slugcomment{Received 2010 December 17; accepted 2011 February 1;
  published 2011 March 10}

\title{Integral Field Spectroscopy of Massive, Kiloparsec-Scale
  Outflows in the Infrared-Luminous QSO Mrk 231}

\author{David S. N. Rupke} \affil{Department of Physics, Rhodes
  College, Memphis, TN 38112} \email{drupke@gmail.com}

\author{and Sylvain Veilleux} \affil{Department of Astronomy,
  University of Maryland, College Park, MD 20742}

\begin{abstract}
  The quasi-stellar object (QSO)/merger Mrk 231 is arguably the
  nearest and best laboratory for studying QSO feedback.  It hosts
  several outflows, including broad-line winds, radio jets, and a
  poorly-understood kpc scale outflow.  In this Letter, we present
  integral field spectroscopy from the Gemini telescope that
  represents the first unambiguous detection of a wide-angle, kpc
  scale outflow from a powerful QSO.  Using neutral gas absorption, we
  show that the nuclear region hosts an outflow with blueshifted
  velocities reaching 1100~\kms, extending $2-3$ kpc from the nucleus
  in all directions in the plane of the sky.  A radio jet impacts the
  outflow north of the nucleus, accelerating it to even higher
  velocities (up to 1400~\kms).  Finally, 3.5~kpc south of the
  nucleus, star formation is simultaneously powering an outflow that
  reaches more modest velocities of only 570~\kms.  Blueshifted
  ionized gas is also detected around the nucleus at lower velocities
  and smaller scales.  The mass and energy flux from the outflow are
  $\ga$2.5 times the star formation rate and $\ga$0.7\%\ of the active
  galactic nucleus luminosity, consistent with negative feedback
  models of QSOs.
\end{abstract}



\section{INTRODUCTION} \label{sec:introduction}

Theory predicts that negative feedback in active galactic nuclei
(AGNs) regulates the growth of supermassive black holes and their
accompanying starbursts \citep{dimatteo05a,narayanan08a,hopkins10a}.
On the scales of galaxies, radio jets power neutral and ionized
outflows with significant mass and energy flux
\citep{morganti05a,morganti07a,holt08a,fu09a,holt10a}.  Wide-angle
outflows on sub-kiloparsec scales are also common in radio-quiet
systems \citep[e.g.,][]{crenshaw05a}.  Precisely how and to what
degree these phenomena are the negative feedback predicted by theory
is an area of active study.

The ultraluminous infrared galaxy (ULIRG) Mrk 231, the nearest
infrared-luminous quasi-stellar object (QSO), is a unique system
because it hosts several types of AGN outflows.  It also combines many
of the phenomena of interest in models of galaxy mergers that then
form powerful QSOs \citep[e.g.,][]{sanders88a,hopkins05a}.  The nuclei
in Mrk 231 have coalesced, but prominent features of an interaction
are still present \citep[e.g.,][]{surace98a}.  It contains a powerful,
unobscured AGN alongside an obscured starburst \citep{veilleux09a}.
Finally, several types of outflows are present: pc and kpc-scale radio
jets \citep{carilli98a,ulvestad99a}; a low ionization, broad
absorption line outflow with blueshifted velocities up to 8000~\kms\
\citep[e.g.,][]{boksenberg77a,boroson91a}, presumably arising at pc
scales; and a kpc scale outflow with velocities exceeding 1000~\kms.

The properties of this third outflow concern us here.  Blueshifted
emission lines led to speculation of its existence
\citep{hamilton87a,krabbe97a}, but emission lines are subject to
misinterpretation, as blueshifted lines could indicate inflow or
outflow \citep[e.g.,][]{veilleux91c}.  Using long slit observations,
\citet{rupke05c} showed conclusive evidence that a $\ga$1000~\kms\
neutral outflow was present at galactocentric radii up to 3 kpc.  More
recently, this outflow was detected in the molecular phase at very
similar velocities, with a marginal detection of extended emission
\citep{fischer10a,feruglio10a}.

The structure and power source of this outflow have remained elusive.
Though the high velocities suggested the influence of the AGN was
helping to power the wind, \citet{rupke05c} were unable to determine
whether the wind was powered by the jet or more widely-directed AGN
energy, and what role the starburst played.  The high molecular gas
velocities observed \citep{fischer10a,feruglio10a} also suggest AGN
influence, but the spatial resolution of these observations made the
power source equally difficult to determine.

In this Letter, we present integral field spectroscopy of the outflow
in Mrk 231, probing both the neutral and ionized phases using \nadl\
interstellar absorption and the \ha\ / \nt\ emission-line complex.
Using sub-kpc spatial resolution, we untangle the structure and
dynamics of the outflow and show that jets, wide-angle AGN influence,
and the starburst all play a role in moving large amounts of gas out
of the central regions of this galaxy.

\section{OBSERVATIONS, REDUCTION, AND ANALYSIS} \label{sec:obs}

Mrk 231 was observed on 2007 July 7 UT with the Integral Field Unit in
the Gemini Multi-Object Spectrograph (GMOS;
\citealt{allington-smith02a,hook04a}) on the Gemini North telescope.
Cloud cover was minimal, and image quality was very high (0\farcs5 -
0\farcs6).  We obtained five exposures of 900 s each, at a position
angle of 35\degr\ to ensure that diffraction spikes from the nuclear
point source were not oriented along the north-south jet axis.  Our
dither pattern centered on the nucleus, with exposures alternating
$\pm$1\farcs5 along one of the field axes.  We used the integral field
unit in two-slit mode and the B600 grating, yielding wavelength
coverage from 5600 to 6950 \AA\ and a spectral resolution of 1.80\AA\
at 6300\AA.

We reduced the data using the IRAF data reduction package provided by
the Gemini Observatory, supplemented with custom IDL routines.  The
0\farcs3 spaxels in the final data cube maximize signal-to-noise ratio
while adequately sampling the seeing disk.  The resulting cube
contains 525 spectra and measures 6\farcs3 $\times$ 7\farcs5, with the
long axis oriented northeast to southwest.

We modeled the continuum and emission lines in each spectrum using
UHSPECFIT, a suite of IDL routines that fits a continuum and emission
lines to spectra \citep{rupke10b}.  We fit the continuum at each point
as a linear combination of the nuclear spectrum and a smooth host
galaxy continuum.  Where possible, we fit two Gaussian velocity
components to the emission lines.  One component represents the
narrow, rotating velocity component, and the second represents an
underlying broad, blueshifted component.

We modeled the \nad\ absorption lines using the method of
\citet{rupke05a}.  One velocity component provided good fits
throughout the cube.  The nearby \ion{He}{1} 5876~\AA\ emission line
was parameterized using the fit to other lines, but in almost all
spaxels this line was absent or very weak.

\begin{figure}
  \includegraphics[width=6.5in]{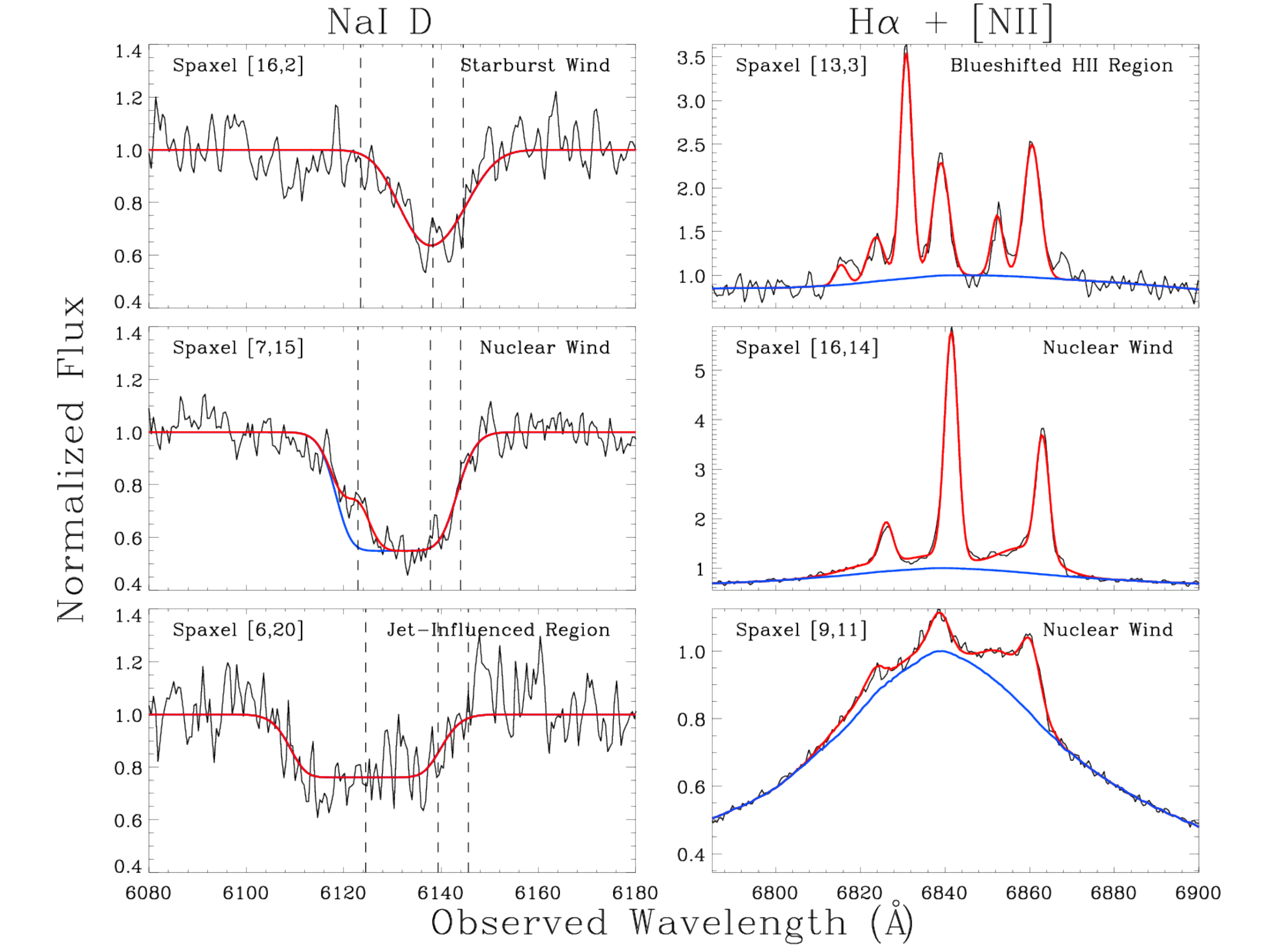}
  \caption{Left: absorption line fits to \nad.  One spectrum is shown
    in each major outflow region (Sections\,\ref{sec:outflow} and
    \ref{sec:discussion}).  The spaxel coordinates of each fit are
    shown in the upper left of each plot; the nucleus is located at
    spaxel [11,13], and spaxels are 0\farcs3 in size.  The vertical
    dashed lines show the locations of \ion{He}{1}~5876~\AA\ and \nad\
    according to the rotation curve.  The red line is the total
    (absorption $+$ emission) fit, while the blue line is absorption
    only.  Right: emission line fits to the \ha\ / \nt\ spectral
    region.  The red line is the total (extended line emission $+$
    broad line region emission $+$ continuum) fit, while the blue line
    is broad line region emission $+$ continuum only.}
  \label{fig:fits}
\end{figure}

Figure \ref{fig:fits} shows six example fits to the emission and
absorption lines.  They illustrate the high quality of the fits.

Our line profile modeling is based on Gaussian velocity profiles.  As
such, we define outflow velocities in this Letter based on the
properties of the normal distribution.  We define negative velocities
to be blueshifted and outflowing.
\begin{align}
  v_{50\%} &\equiv \mathrm{center~of~Gaussian~profile} \notag \\
  & \mathrm{(50\%~of~gas~has~lower~outflow~velocity)}\\
  v_{84\%} &\equiv v_{50\%} - \sigma \notag \\
  & \mathrm{(84\%~of~gas~has~lower~outflow~velocity)}\\
  v_{98\%} &\equiv v_{50\%} - 2\sigma \notag \\
  & \mathrm{(98\%~of~gas~has~lower~outflow~velocity)}
\end{align}

\section{RESULTS} \label{sec:results}

\subsection{Host Galaxy} \label{sec:host-galaxy}

\begin{figure}
  \centering
  \includegraphics[width=5in]{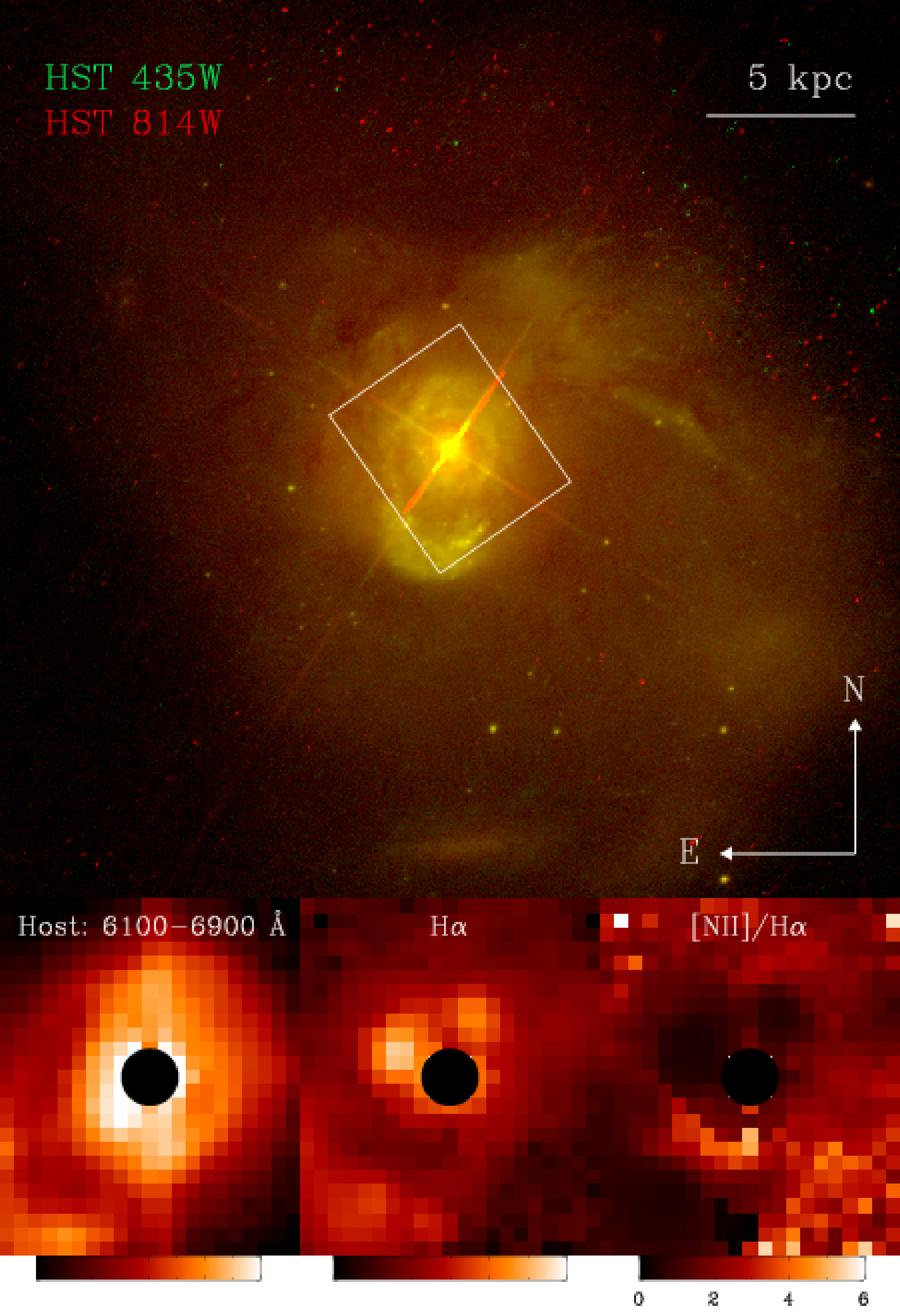}
  \caption{Top: continuum image of Mrk 231 in the 435W and 814W
    filters, from {\it Hubble Space Telescope} observations with the
    Advanced Camera for Surveys.  The field of view of our GMOS data
    (6\farcs3 $\times$ 7\farcs5) is overlaid as a box.  Bottom left:
    host galaxy continuum image in our GMOS field, summed over the
    wavelength range $6100-6900$~\AA, in logarithmic flux units.
    Bottom center: \ha\ emission, in logarithmic flux units.  Bottom
    right: map of the \ntl/\ha\ flux ratio.}
  \label{fig:host}
\end{figure}

Our data reproduce known properties of the host galaxy in Mrk 231.
Figure \ref{fig:host} shows the entire galaxy, with the GMOS field of
view superimposed.  \ha\ emission traces young star formation,
including the edge of a prominent blue arc $\sim$5 kpc south of the
nucleus \citep{surace98a}.  The \nt/\ha\ map reveals high excitation
outside of star forming regions; high extranuclear excitation is
common in ULIRGs \citep{veilleux95a,monrealibero06a}.

The rotation of the central gas disk of Mrk 231 has been modeled using
CO observations \citep{barnes96a,downes98a}, yielding a projected
velocity amplitude of 70~\kms\ along the 90$^\circ$ line of nodes and
a disk inclination of $10^\circ-20^\circ$.  This molecular gas is
concentrated in a $r \sim 1$~kpc disk \citep{downes98a}.  From CO and
\ion{H}{1} observations \citep{downes98a,carilli98a}, we adopt a
systemic velocity of $z = 0.0422$.  This yields a spatial scale of
0.867 kpc arcsec$^{-1}$.

\begin{figure}
  \includegraphics[width=6.5in]{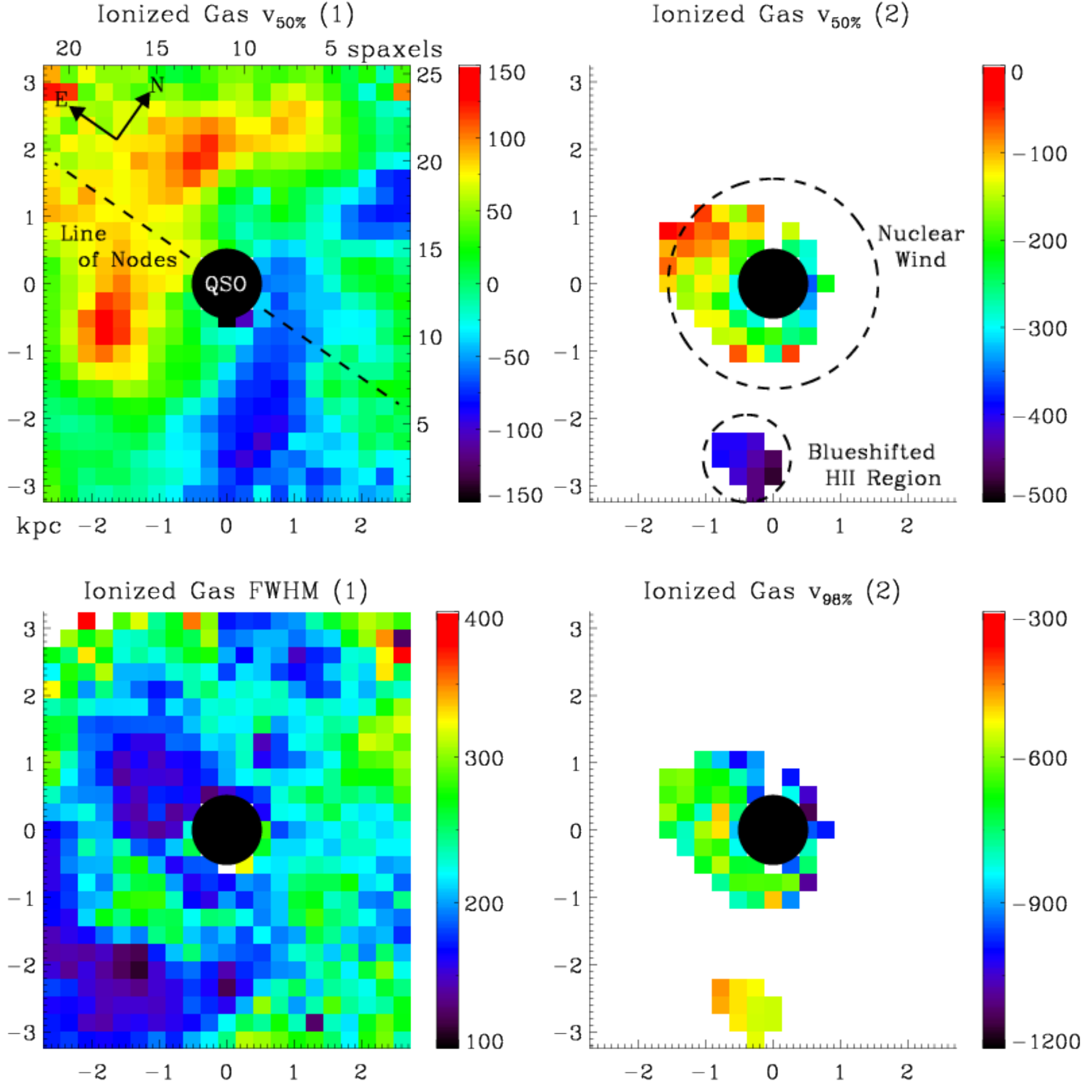}
  \caption{Left: central velocity and FWHM maps of the rotating
    ionized gas component, in \kms.  The ionized gas rotation is
    consistent with that of the nuclear molecular disk.  The line of
    nodes position angle is from \citet{downes98a}. Right: $v_{50\%}$
    and $v_{98\%}$ (Section\,\ref{sec:obs}) for the broad, outflowing
    ionized gas component.  A near-nuclear outflow is present, as well
    as a highly blueshifted \htwo\ region in the southwest.}
  \label{fig:emission}
\end{figure}

The rotation of the ionized gas lines up with that of the molecular
component, despite the difference in extinction of the two components
(Figure \ref{fig:emission}).  Our data trace this rotation to larger
radii than the CO observations.  Deviations from the typical galaxy
rotation curve are evident, but discussion of these features is
outside the scope of this Letter.

\subsection{Outflow} \label{sec:outflow}

The second ionized gas component is blueshifted and much broader than
the rotating component.  Within 1.5~kpc of the nucleus, the center
velocity of the second component averages $\langle v_{50\%} \rangle =
-150$~\kms, with larger velocities closer to the nucleus.  Because
these components are also very broad, with $\langle$FWHM$\rangle =
700$~\kms, the ionized gas reaches $\langle v_{98\%} \rangle =
-760$~\kms.  The region of broad, blueshifted emission is asymmetric,
with higher velocities to the north and more extended emission to the
east.

\begin{figure}
  \includegraphics[width=6.5in]{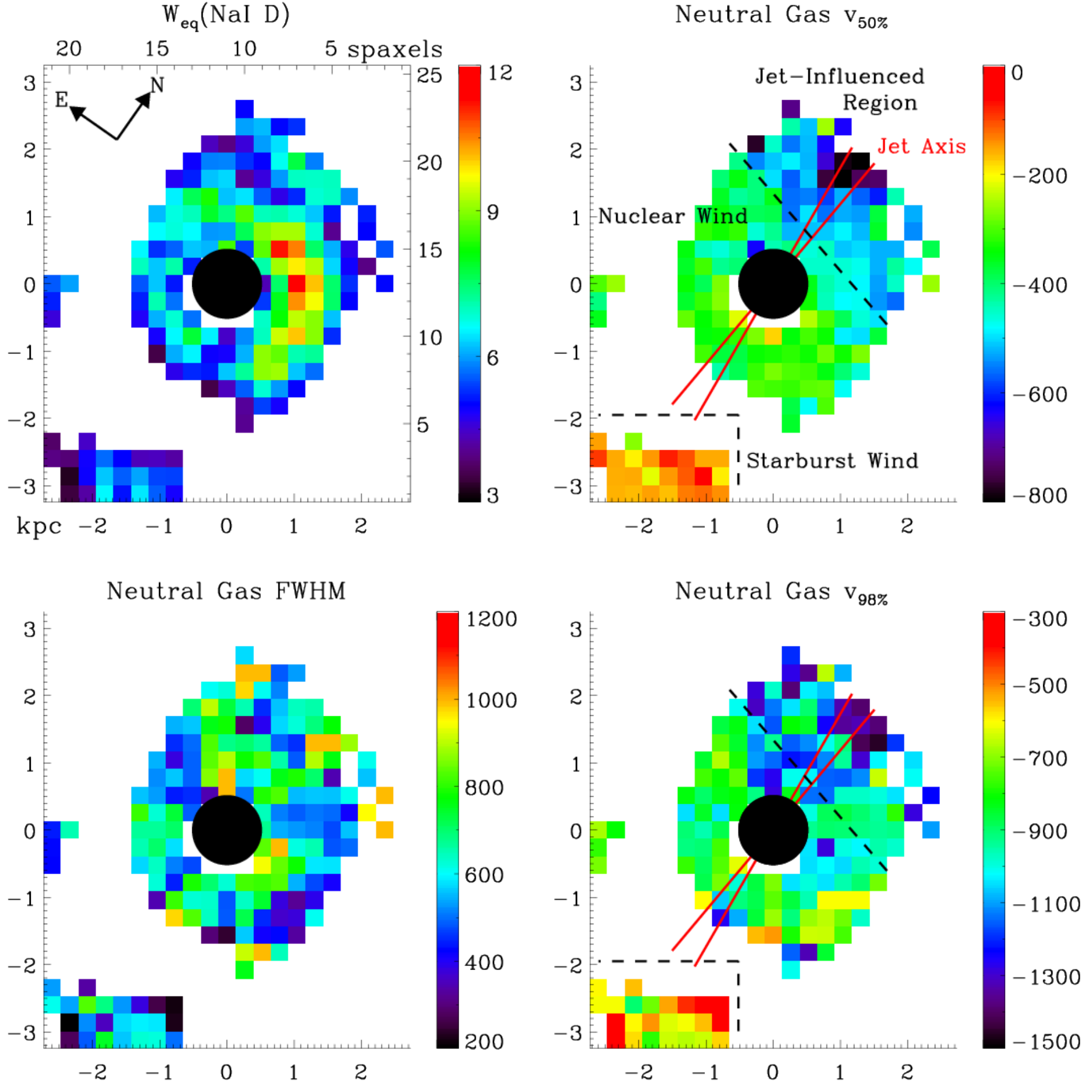}
  \caption{Equivalent width, central velocity, FWHM, and $v_{98\%}$
    maps of \nad.  A nuclear outflow extends from the nucleus up to
    $2-3$~kpc in all directions (as projected in the plane of the
    sky).  The high velocities suggest that the AGN powers the nuclear
    wind.  The northern quadrant of the nuclear wind is further
    accelerated by the radio jet.  A lower-velocity starburst-driven
    outflow is present in the south.}
  \label{fig:abs}
\end{figure}

The neutral atomic gas in Mrk 231, as traced by \nad, is strikingly
different from the ionized gas (Figure \ref{fig:abs}).  At the
4$\sigma$ level, \nad\ absorption extends from radii of 0.5 to 3 kpc,
which is farther than the ionized outflow.  (Closer than 0.5~kpc, the
nuclear emission washes out the signature of the host galaxy.)  A
significant area of absorption is also seen atop the blue continuum
peak 3.5 kpc south of the nucleus.

The neutral gas velocities are also much higher than the ionized gas
velocities.  The \nad\ velocity maps consist of three distinct
regions, whose velocity averages are found in Table \ref{tab:vels}.
The lowest velocities lie atop the southern star-forming arc.  Higher
velocity, broader components are found in the outflow surrounding the
nucleus (not including the northern quadrant).  The highest velocities
are found in the northern quadrant of the nuclear outflow.  In those
parts of the northern quadrant that line up with the radio jet in
Mrk~231, the velocities reach $-$1400~\kms.

The velocities observed in the arc are comparable to those measured in
starburst-dominated ULIRGs (Table \ref{tab:vels};
\citealt{martin05a,rupke05b}).  However, those in the nuclear wind are
significantly higher; they are more like those found in some Seyfert
ULIRGs \citep{rupke05c,krug10a}.

\section{DISCUSSION} \label{sec:discussion}

ULIRGs host powerful starburst-driven outflows
\citep{heckman90a,heckman00a} that arise in all ULIRGs
\citep{rupke05b} and show kpc scale extents \citep{martin06a,shih10a}.
They have $\langle v_{98\%} \rangle = -450$~\kms\ in systems whose
infrared luminosity is dominated by star formation
\citep{rupke05b,martin05a}.  \citet{rupke05b} calculate that ULIRG
mass outflow rates are 20\%\ of the star formation rate on average.
However, evidence has remained elusive for large-scale outflows that
are clearly AGN-driven in LIRGs or ULIRGs \citep{rupke05c}.

The data we present here are clear: there is a neutral,
$\sim$1000~\kms\ outflow in Mrk 231 that extends in every direction
from the nucleus (as projected into the plane of the sky) out to at
least 3 kpc.  Such high velocities have not been seen in starburst
ULIRGs (Table \ref{tab:vels}), providing strong circumstantial
evidence that this wide-angle nuclear wind is driven by radiation or
mechanical energy from the AGN.

From what we know about the structure of galactic winds
\citep{veilleux05a}, the molecular disk in Mrk 231 collimates this
nuclear wind.  Given the disk's almost face-on orientation, we must be
looking ``down the barrel'' of a biconical outflow.  The other end of
this bicone is receding from us, behind the galaxy disk and therefore
invisible at optical wavelengths.

It is apparent from the velocity map that the north$-$south radio jet
in Mrk 231 \citep{carilli98a,ulvestad99a} is coupling to the nuclear
wind, accelerating the neutral gas to even higher velocities.  The jet
is not constrained to emerge perpendicular to the disk, and thus
produces an asymmetric effect.  The present data imply that the
northern arm of the large-scale jet is on the near side of the
molecular disk.  Neutral outflows driven by jet interactions with the
interstellar medium (ISM) on kpc scales have also been observed in
radio galaxies in \ion{H}{1} absorption
\citep{morganti05a,morganti07a}.  The newly discovered jet-wind
interaction in Mrk 231 appears to be similar, though this time the jet
accelerates an already in situ wind.

\citet{carilli98a} and \citet{taylor99a} studied the diffuse radio
continuum emission from Mrk~231, which is symmetric about the nucleus
on scales of 100~pc to 1~kpc.  They hypothesized that this emission is
produced by in situ electron acceleration, but could not rule out that
the AGN distributes these electrons through a wide-angle outflow.  Our
data are further evidence for the in situ interpretation, since we now
know that the AGN outflow reaches larger scales and is asymmetrically
accelerated.

Along with these AGN-driven outflows, Mrk 231 also hosts a
starburst-driven wind.  Blueshifted velocities with $\langle v_{98\%}
\rangle = -570$~\kms\ are observed in the star-forming arc south of
the nucleus.  These are comparable to the maximum velocities observed
in other starburst ULIRGs (Table \ref{tab:vels}).  We cannot rule out
that the underlying continuum could be backlighting the nuclear wind
at this location.  However, a simpler explanation is that in situ star
formation is driving an off-nuclear wind.

The ionized outflow in Mrk 231 overlaps spatially with the neutral
outflow near the nucleus and shows broad velocity profiles.  It also
shows higher velocities to the north.  However, the velocities and
extent of this wind phase are more modest.  The ionized outflow in Mrk
231 could be another phase of the nuclear wind, providing further
evidence that the different gas phases in ULIRG winds are not strongly
coupled \citep{rupke05c,shih10a}.  It may also be the classic AGN
narrow line region \citep[e.g.,][]{crenshaw05a}.

The nuclear wind is clearly the one detected previously in the neutral
\citep{rupke05b} and molecular \citep{fischer10a,feruglio10a} phases.
\citet{rupke05b} found an extended wind (on kpc scales), but only had
data in a north-south slice.  \citet{feruglio10a} marginally resolved
the molecular component out to radii of $\sim$0.6~kpc, and estimated
molecular mass and energy fluxes of several hundred \smpy\ and
$\sim$10$^{44}$ erg s$^{-1}$.

The current data are a significant improvement on these results.  They
clearly resolve the structure of the wind on sub-kpc scales and reveal
that more than one power source is acting.  To produce a wide-angle
outflow, the AGN energy and momentum must be injected into the ISM
over a substantial volume through mechanical or radiative processes.
Either (1) wide-angle winds from the AGN accretion disk or broad line
region or (2) coupling of the AGN radiation to the dust in the wind
are possible explanations \citep[e.g.,][]{crenshaw03a,hopkins10a}.
However, the high-velocity component of the wind that lies north of
the nucleus must experience an additional, asymmetric force.  The
north-south jet in Mrk 231 is a natural explanation.

Using a simple model, we can estimate the mass and energy fluxes
($\dot{M}$ and $\dot{E}$) in the nuclear wind \citep{rupke05b,shih10a}.  We
assume a face-on wind for simplicity.  We model the wind as emanating
from the nucleus in a thin shell, and compute a time-averaged mass
flux.  \citet{rupke05b} also discuss a thick-shell model.  However,
computing the mass outflow rate for this model would require
integrating hypothetical velocity profiles over each line of sight;
such an exercise is beyond the scope of this Letter and would increase
the number of unconstrained parameters.

There are two significant unknowns in the thin shell model.  The first
is the radius of the shell.  Given the spatial smoothness of the \nad\
velocity map, it must be larger than 2~kpc, or we would observe the
velocity to decrease (in projection) with increasing radius.  However,
it cannot be too large, or we would observe a more extended absorption
region.  We use a fiducial radius of 3~kpc; larger values will
increase the resulting $\dot{M}$ and $\dot{E}$.  The second
uncertainty is the ionization state of Na.  As in previous work
\citep[e.g.,][]{rupke05a}, we use Milky Way measurements as a baseline
and assume $N$(\ion{Na}{1})/$N$(Na) $= 1-y = 0.1$, where $y$ is the
ionization fraction.  The fluxes are inversely proportional to
$(1-y)$, such that a more ionized wind will yield higher $\dot{M}$ and
$\dot{E}$.

With this model, we compute $\dot{M}$ $=$ 420 ($R$/3~kpc)
$[0.1/(1-y)]$ \smpy\ and $\dot{E}$ $=$ $7.3\times10^{43}$ ($R$/3~kpc)
$[0.1/(1-y)]$ erg s$^{-1}$.  For comparison, the star formation rate
and AGN luminosity in Mrk 231 are 172 \smpy\ and $1.1\times10^{46}$
erg s$^{-1}$.  (These numbers are computed by assuming the bulk of the
galaxy's radiation emerges in the far-infrared, using the AGN
contribution to Mrk~231's bolometric luminosity from
\citealt{veilleux09a}, and calculating the star formation rate with
the formula from \citealt{kennicutt98a}.)  The wind is clearly
removing significant amounts of gas from the nucleus -- at a level of
2.5 times the star formation rate.  This mass outflow rate is also
strikingly similar to that estimated by \citet{feruglio10a} from a
different tracer.

The energy outflow rate is about 0.7\%\ of the radiative luminosity of
the AGN, meaning that only a small part of the radiative output of the
AGN has to couple to the outflow.  This is remarkably similar to the
required coupling efficiency needed for an AGN wind in the two-phase
model of \citet{hopkins10a}.  In this model, the AGN couples to the
hot ISM, driving a diffuse outflow.  This diffuse outflow disrupts
cold clouds, which are propelled outward by the diffuse wind and by
radiation pressure from the AGN.  Together, the hot and cold winds
provide negative feedback on the AGN and star formation.

\section{SUMMARY} \label{sec:summary}

The galaxy Mrk 231 is in the late stages of a major merger, hosts a
QSO and an obscured starburst, and exhibits gas outflows in multiple
forms.  We show in this Letter that there is a massive, energetic, and
neutral outflow extending in all directions from the nucleus (as
projected in the plane of the sky) to radii of at least 3 kpc.  This
outflow has blueshifted velocities that reach 1100~\kms.  On the
axis of the radio jet in Mrk~231, $v_{98\%}$ is as large as 1400~\kms.
Mass and energy outflow rates are $\ga$2.5$\times$ the star formation
rate and $\ga$0.7\%\ of the AGN luminosity.  The former suggests
strong negative feedback to star formation, and the latter is
consistent with the coupling efficiency required in the AGN feedback
model of \citet{hopkins10a}.

By comparing the velocities in this wind to those in the average
starburst-driven ULIRG, we conclude that this wide-angle wind is
likely powered by the central AGN, and its power source must be
directed fairly symmetrically.  The northern quadrant of this nuclear
outflow experiences additional acceleration due to the radio jet in
Mrk~231.  Furthermore, a starburst-powered wind arises in the extended
galaxy disk to the south of the nucleus.

These data illuminate the structure and power source of the kpc scale
wind in Mrk 231, and show that strong feedback occurs as a result.
The ubiquity of winds in major mergers is well-known, but our
understanding of their full impact on galaxy evolution is improving
through integral field observations (\citealt{shih10a} and the present
work), Herschel studies \citep{fischer10a,sturm10a}, and ground-based
molecular observations \citep{feruglio10a}.

\acknowledgments This Letter was based on observations obtained at the
Gemini Observatory, which is operated by AURA under a cooperative
agreement with NSF on behalf of the Gemini partnership.  It also made
use of NASA/ESA {\it Hubble Space Telescope} data obtained from the
Hubble Legacy Archive.  S.V. acknowledges partial support by the
National Science Foundation through AST/EXC grants AST0606932 and
AST1009583.

\def\eprinttmppp@#1arXiv:@{#1}
\providecommand{\arxivlink[1]}{\href{http://arxiv.org/abs/#1}{arXiv:#1}}
\def\eprinttmp@#1arXiv:#2 [#3]#4@{\ifthenelse{\equal{#3}{x}}{\ifthenelse{
\equal{#1}{}}{\arxivlink{\eprinttmppp@#2@}}{\arxivlink{#1}}}{\arxivlink{#2}
  [#3]}}
\providecommand{\eprintlink}[1]{\eprinttmp@#1arXiv: [x]@}
\providecommand{\eprint}[1]{\eprintlink{#1}}
\providecommand{\adsurl}[1]{\href{#1}{ADS}}
\renewcommand{\bibinfo}[2]{\ifthenelse{\equal{#1}{isbn}}{\href{http://cosmologist.info/ISBN/#2}{#2}}{#2}}

\clearpage

\begin{deluxetable}{lccccr}
  \tablecaption{Outflow Velocities in Mrk 231\label{tab:vels}}
  \tablewidth{0pt}

  \tablehead{ \colhead{Label} & \colhead{Phase} &
    \colhead{$\langle$FWHM$\rangle$} & \colhead{$\langle v_{50\%}
      \rangle$} & \colhead{$\langle v_{84\%} \rangle$} &
    \colhead{$\langle v_{98\%} \rangle$}\\
    \colhead{(1)} & \colhead{(2)} & \colhead{(3)} & \colhead{(4)} &
    \colhead{(5)} & \colhead{(6)} }

  \startdata
  Mrk 231 starburst wind & Neutral & 510 & -150 & -360 & -570 \\
  Mrk 231 nuclear wind   & Neutral & 600 & -360 & -620 & -880 \\
  \nodata  & Ionized\tablenotemark{a} & 690 & -160 & -450 & -750 \\
  Mrk 231 jet-influenced region & Neutral & 640 & -520 & -800 & -1070 \\
  \hline
  Starburst ULIRGs\tablenotemark{b} & Neutral & 330 & -170 & -310 & -450
  \enddata

  \tablecomments{Column 2: gas phase.  Column 3: full width at half
    maximum of velocity distribution, in \kms.  Column 4: velocity at
    the center of (Gaussian) velocity distribution.  Column 5:
    $v_{84\%} \equiv v_{50\%} - \sigma$; 84\%\ of the outflowing gas
    has velocities closer to 0.  Column 6: $v_{98\%} \equiv v_{50\%} -
    2\sigma$.}

  \tablenotetext{a}{For the broad, outflowing component.}
  \tablenotetext{b}{Computed from starburst-driven ULIRGs in
    \citet{rupke05b}.  This is a sample average, as opposed to the
    spatial average for Mrk~231.}

\end{deluxetable}

\end{document}